\documentclass[prb,twocolumn]{revtex4}
\usepackage{graphicx}
\begin{document}
\title{Condensation of the atomic relaxation vibrations in lead-magnesium-niobate at $T=T^*$}
\author{Sergey Prosandeev$^{1,2}$, Igor P. Raevski$^{2}$, Maria A.
Malitskaya$^{2}$, Svetlana I. Raevskaya$^{2}$, Haydn Chen$^3$, Chen-Chia Chou$^4$,
 and Brahim Dkhil$^{5}$}
\affiliation{
$^1$ Physics Department and Institute for Nanoscience and Engineering,
University of Arkansas, Fayetteville, AR 72701, USA \\
$^{2}$Department of Physics, Research Institute of Physics,
Southern Federal University, Rostov on Don 344090, Russia \\
$^3$University of Macau, Av. Padre Tomas Pereira, Taipa, Macau,
Republic of China \\
$^4$National Taiwan University of Science and Technology,
43 Keelung Road, Sec.4.  Taipei, Taiwan 106, Republic of China \\
$^{5}$Laboratoire Structures, Proprietes et Modelisation des Solides,
Ecole Centrale Paris, UMR-CNRS 8580, 92290 Chatenay-Malabry, France}

\begin{abstract}
We present neutron diffraction, dielectric permittivity and photoconductivity measurements, evidencing that lead-magnesium niobate
experiences a diffuse phase transformation between the spherical glass and quadrupole glass phases,
in the temperature interval between 400 K
and 500 K, with the quadrupole phase possessing extremely high magnitudes of dielectric permittivity. Our analysis shows that the integral diffuse scattering intensity may serve as an order parameter for this transformation. Our experimental dielectric permittivity
data support this choice. These data are important for the aplications desiring giant dielectric responses, in a wide temperature intervals and not related to electron's excitations.
\end{abstract}
\maketitle

\section{Introduction}
Lead magnesium niobate PbMg$_{1/3}$Nb$_{2/3}$O$_{3}$ (PMN) is a
canonical relaxor material characterized by an extremely large
magnitude of dielectric permittivity over a broad temperature
range and having logarithmically-wide frequency dispersion
\cite{Smolenskiy,Cross,Viehland} that is strongly demanded in many valuable
applications. For example, this is important to produce small-size
capacitors robust to the temperature change. Dilution of PMN with
ferroelectrics led to the discovery of one of the best
piezoelectrics ever found, which have now extremely wide
applications, in all brunches of life, science and technology
\cite{Zhang}. At the same time, the understanding of the main
property of relaxors, which is the colossal dielectric response,
is still debated \cite{Glazunov,Isupov,Colla,Vugmeister-Glinchuk,Blinc1,Blinc2,Blinc3,Kalinin,Cowley,Bokov}.
The source for the main properties of relaxors is usually related
to the temperature growth of the polar nano regions (PNR's) on
cooling \cite{Isupov}. It is believed that PNR's start their
development, at a temperature called the Burns temperature of
$T_d=620\,\rm{K}$, and increase their strength on cooling
\cite{Burns}.

Chu, Setter and Tagantsev \cite{Tagantsev} suggested that $T_d$ might manifest a local phase transformation and heterophase fluctuations. According to Ref. [\onlinecite{Okuneva}], this phase is characterized by a critical change of the lower boundary of the Pb off-centering magnitude. Specifically, this lower boundary, above $T_d$, is zero, but critically increases on cooling below $T_d$.  We will call the phase below this temperature as Spherical Glass, because the Pb-off-centerings were shown to be spherical \cite{Vakhrushev}. The spherical model was recently confirmed also in Ref. [\onlinecite{Jirka}], and it is also in line with the suggestion that this phase corresponds to the Heisenberg-like universality class, while at lower temperatures reduces to the Ising universality class \cite{Stock}. In Ref. [\onlinecite{Bussmann}] the emergence of the relaxor properties at $T=T_d$ was related to the crossover between the lattice and relaxational dynamics, leading to a specific phonon confinement called also breathers.

We assume that this phase freezes only the magnitude, but not the direction of the Pb displacement. This assumption implies that the corresponding order parameter is scalar. The correlation of these scalar (looking like a Higgs boson \cite{Higgs}) off-centerings can happen via the lattice volume modulation, which is scalar too. It is interesting to notice also that some of such off-centerings can have intense rotational dynamics, and the transverse instability of such rotators might result in specific nontrivial relaxations \cite{Bokov} and contributions to dielectric permittivity \cite{Langevin} that, actually, has been observed as a non-Landau contribution \cite{nonlinear}.

Vogel-Fulcher freezing temperature $T_f$ is another famous temperature in relaxors  \cite{Viehland,Colla}. Below this temperature, PMN reaches a glass-like state associated with  critical slowing down of the local dipoles on cooling together with gradual broadening of the relaxation frequency spectrum \cite{Tagantsev-Vogel}. Nevertheless, in contrast to the glass-like idea \cite{Colla}, $T_f$ can be alternatively considered as the temperature of the emergence of the polarization nanodomains, controlled by the quenched random electric fields \cite{RandomField,domains} arising from the $Mg^{2+}$ and $Nb^{5+}$ random distribution over the crystal lattice. A combined concept has been introduced by Glinchuk, according to which  the low temperature phase in PMN can be called as Mixed Ferroglass \cite{Glinchuk}. This author defines two quantities, $E_0$ is the most probable magnitude of the long-range dipolar field and $\Delta E$ is the half width of the random field distribution function. When $\Delta E$ is much larger than $E_0$, a glass order stabilizes. If $\Delta E$ is much smaller then ferroelectricity shows up. In between of these two limits, the Ferroglass is the stable phase, which is a mixture of the glass and ferroelectric phases. The formation of the Ferroglass on cooling has been thoroughly studied directly by neutron diffraction \cite{Egami} and dielectric spectroscopy \cite{Colla} methods.

Knowing the observations at both $T_d$ and $T_f$ peculiar temperatures, several elegant theoretical models have been proposed. They include, for instance, the Quenched-Random-Field-Model \cite{RandomField}, the Random-Site-Electric-Dipole-System theory \cite{Vugmeister-Glinchuk,Glinchuk-Stephanovich} or the most recent and accepted Spherical-Random-Bond--Random-Field Model \cite{Blinc1,Blinc2,Blinc3}. This latter model considers a new type of glass by introducing an analog of the Edvards-Anderson parameter in the theory of relaxors.

It is worth also mentioning first-principles calculations of relaxors \cite{Barton}, which exploited the idea \cite{Akbas} that the ferroelectric relaxors possess nanoscale chemically ordered regions facilitating the emergence of PNR's. In contrast to this idea, some other calculations have shown a possibility to calculate some relaxor properties without the use of the idea about the chemically ordered regions \cite{Laurent-BZT,Iniguez,Rappe}. The idea of the frustration of ferroelectricity has been also considered in the frame of recent first-principles calculations \cite{Nakhamson,Narayani}.

In addition to the relaxor-like features related to $T_d$ and $T_f$,
Viehland \textit{et al} emphasized the fundamental
importance of temperature $T_c=400\,\rm{K}$, which could manifest a new phase having glass features \cite{Viehland-Curie}. Interestingly, the authors got this temperature by extrapolating the Curie-Weiss temperature, from the high-temperature part of dielectric permittivity. Dkhil \textit{et al} \cite{Brahim} found a critical temperature $T_L \approx 400 K$, on the basis of the lattice parameter temperature behavior and X-Ray and neutron integrated diffraction intensity.
In qualitative agreement with these
findings, Svitelskiy \textit{et al} \cite{Svitelskiy} found that E-Raman-lines split up, critically, at $T^*=350K$, on cooling, with the ratio of the components of $2:1$. The authors explained this splitting by a (local) tetragonal distortion.
Recent measurements \cite{Dulkin} of the thermally stimulated acoustic emission revealed another temperature,  $T_1^*=500\,\rm{K}$, which is significantly higher than 400 K. At this temperature, the acoustic emission had a pulse, on cooling. These authors related this pulse to the formation of static PNR's.
Most recently, new measurements
\cite{Gehring} of diffuse neutron scattering, with a higher than earlier precision \cite{footnote1}, revealed the former temperature,
$T_2^*=400\,\rm{K}$, again. According to the authors, this temperature marks the start of the diffuse neutron scattering, having a
so-called butterfly shape, on cooling. Less precise earlier measurements exhibited a strong tail in this integral intensity \cite{Brahim}. The authors \cite{Gehring} concluded that PNR's develop on cooling starting from $T_2^*=400\,\rm{K}$ rather than from $T_d$.
To the best of
our knowledge, all these facts have not got any reasonable
explanation, in the frame of a microscopic model of relaxors, so
far. Actually, this transformation has not been taken into account in the previous theoretical models \cite{RandomField, Vugmeister-Glinchuk,Blinc1,Blinc2,Blinc3}. In the present paper, we will focus on the significance of this issue. (Notice that the notations $T_1^*$ and $T_2^*$ are ours. The authors of Refs. [\onlinecite{Dulkin,Gehring}] both used the notation $T^*$ as in Ref. [\onlinecite{Svitelskiy}]).

We will relate all these $T^*$'s, $T_c$, and $T_L$ to different stages of the freezing of the random local tetragonal distortions, on cooling, and this is why we will call this phase as quadrupole glass (see also studies of other quadrupole glasses in Refs. [\onlinecite{RandomField,Ivliev,Vollmayr,Michel,Pirc}]. This phase differs from the mixed ferroglass phase by the fact that the local dipoles are still dynamic, but being compared with the spherical glass, they are now strongly cooperative, not only in the magnitude but also in the direction. We will argue, in the manuscript, that the quadrupole glass phase transition is diffuse, and this makes the dielectric permittivity also diffuse. Let us remind that this diffusness has been considered so far as one of the main features of the relaxors \cite{Smolenskiy,Cross,Viehland}. Now, we are going to give more details, derivations, and experiments, supporting this picture of PMN.

The plan of the paper is the following. Section 2 discusses the temperature dependence of the lattice parameter in PMN. Section 3 suggests the idea of the quadrupole glass. Section 4 reviews the diffuse scattering of neutrons in PMN. Section 5 addresses the main point of the paper, the giant dielectric response in PMN. Section 6 gives some scenarios of the possible local tetragonal distortion, in the temperature interval of interest. Section 7 presents our data of the photoelectric current. Finally, Section 8 summarizes our results and discusses possible implications.

\section{Temperature dependence of the lattice parameter in PMN}
In order to illustrate an experimental evidence of the quadrupole phase, first, we present the temperature dependence of the lattice
parameter $a$ in PMN \cite{footnote2}. These data manifest two different anomalies, at $T_1^*$ and
$T_2^*$ (the anomaly at $T_2^*$ was denoted earlier also as $T_c$
\cite{Dulkin}). At $T_1^*$, the lattice parameter temperature dependence has the
strongest curvature. Contrary to this, at $T_2^*$, the temperature
change of the lattice parameter vanishes, at all.

Let us discuss,
first, the change at $T_1$*. According to general thermodynamics
\cite{Landau}, the largest curvature of $a(T)$ results in the
largest thermally stimulated strain. Indeed,
\begin{equation}
\label{eq1}
\Delta \eta =\frac{1}{V}\left[ {\left( {\frac{dV}{dT}} \right)_{p} +T\left(
{\frac{d^{2}V}{dT^{2}}} \right)_{p} } \right]\Delta T
\end{equation}
$\Delta T$ is the temperature lowering, $V$ is the volume of the sample,
and $\Delta \eta $ is the change of the strain on temperature cooling. The former term in
eq. (\ref{eq1}) corresponds to the straight dependence of the
volume on temperature. The latter one originates from the
dependence of the heat capacity on pressure \cite{Landau}:

\begin{equation}
\left ( \frac{dc_{p}}{dp}\right)_{T} =-\left ( \frac{d^{2}V}{dT^{2}} \right)_{p}
\end{equation}
We believe \cite{Dulkin09} that this last term in equation (1) is responsible for
the strong thermally stimulated acoustic emission at $T=T_1$*. Indeed, the largest curvature of the lattice parameter temperature change implies the largest second derivative of the volume with respect to temperature, and this results, according to equation (\ref{eq1}), in the largest strain, which can trigger the thermally stimulated acoustic emission (via, for example, appearance of cracks).

The significance of temperature T$_2$* deserves special attention.
At this temperature, the thermal change of the lattice parameter
disappears and, correspondingly, the lattice anharmonicity
vanishes. This kind of transformation has been found also in other
relaxors \cite{Dulkin,BZT-a}.

For the sake of illustration, we want to make some bridge between this phenomenon and the temperature dependency of the density of $^4$He, at so called $\lambda$ point. $^4$He has a discontinuity at this point, above which the density is rapidly growing on cooling, while, below, it saturates after showing some small spike \cite{Tipler}. This fact was
explained by a phase transformation, at the $\lambda$ point, from a classic liquid regime
to quantum one. In the quantum He II phase, below the $\lambda$ point, the $^4$He bosons
cooperatively and synchronically rotate, in a vortex. Thus, the additional cooperative quantum permanent movement of the bosons makes the density of the matter to saturate.

Quantum paraelectrics are other examples, which possess extremely high values of dielectric susceptibility in a wide temperature interval owing to quantum beatings \cite{Muller}. The dielectric permittivity in the quantum paraelectrics saturates, below the so called saturation temperature (which is typically of 50 K).

We emphasize here this kinetic aspect, because dielectric permittivity in the temperature interval below $T=T_2^*$ and above $T=T_f$, in PMN, possesses giant magnitudes, which can be explained by the extreme softness of the vibration or/and relaxation modes. Being based on these analogies, one can expect that, in PMN, at $T=T_2$*, there also
appears a cooperative atomic movement (not necessarily totally
synchronized quantum mechanically), which breaks the phonon
anharmonicity.

It is worth noting that the saturation of the dielectric permittivity in relaxors, below some temperature, has been already discussed \cite{Vug-Rab,Kutnjak,Pirc-ECE}. For example, in Ref. [\onlinecite{Vug-Rab}], the dielectric permittivity is represented in the form
\begin{equation}\label{epsilon}
 \varepsilon=\frac{\varepsilon_0}{1-\kappa f(\omega,T)}
\end{equation}
where $\varepsilon_0$ is constant,
\begin{equation}
f(\omega,T)=\left < \frac{1}{1+i\omega \tau} \right >
\end{equation}
with the relaxation time given by the Vogel-Fulcher formula
\begin{equation}
\tau=\tau_0 \exp \left (\frac{U}{T-T_f} \right )
\end{equation}
Here $\tau_0$ and $U$ are constants. $\kappa$ in equation (\ref{epsilon}) is given by the Landau-type expansion
\begin{equation}
\kappa=a_1+3a_3P_s^2+5a_5P_s^5
\end{equation}
where $P_s$ is the spontaneous polarization, and the coefficient $a_1$ is given by a Barrett-like quantum formula

\begin{equation}
a_1 = 0.95 T_s \tanh \left ( \frac{0.5T_s}{T} \right )
\end{equation}
Here $T_s$ is a characteristic temperature, which can be called also the saturation temperature (in the spirit of the Barrett formula \cite{Muller}). The theory under discussion does not give a clue about the physical origin of this temperature. Notice that only dc dielectric susceptibility saturates, at low T, below $T_s$, while, at any finite frequency, formula (\ref{epsilon}) gives a one-peak function, with the peak position dependent on the frequency.

In the same year, experimental data, obtained for PMN, confirmed this kind of saturation of the dc dielectric susceptibility \cite{Kutnjak}. Later on \cite{Blinc1,Vug-glass,Blinc4,Pirc-ECE}, it was realised that the dielectric susceptibility can be presented in the form of the Sherrington-Kirkpatrick formula \cite{Binder}

\begin{equation}
\chi=\chi_0+\frac{C(1-q)}{T-\theta (1-q)}
\end{equation}
where $C$, $\chi_0$ and $\theta$ are constants, and $q$ is the Ferroglass Edwards-Anderson parameter \cite{E-A}. Thus, the deviation of the temperature evolution of the dielectric permittivity from the Curie-Weiss law was related to the emergence of the glass order parameter at $T=T_f$ \cite{foot-vug}. In the next sections, we will show that this understanding is not complete and requires to be supplemented by another temperature, namely, $T^*$.

\section{Quadrupole glass idea}

Besides the lattice parameter measurements and thermally stimulated acoustic emission described above, there are other evidences of the peciliar behavior of PMN in the close vicinity of $T=T^*$, which we have already mentioned in Introduction. These are dielectric permittivity anomaly \cite{Viehland-Curie}, which was described as the emergence of a glass behavior, around 400 K, and the X-Ray and neutron integrated diffraction intensity \cite{Brahim,Gehring}, which show clear appearance of the integral intensity around 400 K. We will treat both these evidences below, with the help of a model describing the quadrupole glass formation. At last, but not least, Raman spectra in many studies (see e.g. Ref. [\onlinecite{PMN}])  confirmed to have the tetragonal-like splitting found, first, in  Ref. [\onlinecite{Svitelskiy}].
In this section, we suggest a model of the transformation of the spherical glass into the quadrupole glass. Then, we will fit this model to the experimental data. Raman spectra will be discussed in Sec. 6.

Let us focus on the lead atomic displacements in PMN.
According to experimental findings, below
$T=T_d$, the lead ions get off-center, and the strength of these
off-centering's increases on cooling \cite{Okuneva}. This conclusion follows from the fact that the radial distribution of the lead displacements, below $T=T_d$, gets split into two sets, with a gap in between. The gap appears, first, at those lead positions, which were not disturbed by the random fields (high-temperature lead displacement is absent, for such positions).

These off-centerings being combined into polar nano regions (PNR's) increase their strength on cooling, and, thus, increase the strength of the PNR's. However, this increase has some limit, because, when the random PNR's start touching each other, they organize larger superclusters (not necesarily having global dipole moments, because of the disorder of the dipole directions in each of the PNR's), and, finally, these superclusters organize a net through the whole crystal. For the sake of illustration, let us consider a simple model free energy describing two interacting PNR's:

\begin{equation}
\label{eq2}
F=\frac{A}{\left( {r_{0} -2r} \right)^{\nu}}+\alpha \left( T \right)v+\beta
v^{2}
\end{equation}
where $r_{\mathrm{0}}$ is the distance between the lead pairs, $r$
is the correlation radius, the power $\nu$ is supposed to be
large compared to 1, $v(T)$ is the relative correlation volume (or density of the spherical off-centerings). $\alpha
\left( T \right)$ is the inverse local (spherical) susceptibility,
which decreases on cooling and acquires negative values below
$T_d$, resulting in the appearance of finite off-center spherical
displacements. The first term in equation (\ref{eq2}) represents
the repulsion term which increases the free energy on approaching
the correlation spheres of the nearest PNR's, each to
the other, and has, originally, elastic nature \cite{KLT}. The
next two terms ($\beta $ is supposed to be positive, for
simplicity) define the temperature dependence of the volume $v$,
in the case, when the spherical centers are far each from the
other. Finally, by minimizing the free energy (\ref{eq2}), one can
find the equilibrium radii of the nearest PNR's. The average radius can be given by the cubic root of the inverse concentration of the lead pairs.

The emergence of the quadrupole glass
can be determined by the magnitude of the
corresponding quadrupole glass order parameter, which can be
constructed in the same way as the Edwards-Anderson parameter
\cite{Laurent-BZT,E-A}. For this purpose, one should take into
account that this parameter is conjugate to the corresponding
random strain variance, which makes the phase transition
spread out over a significant temperature range, as we
shall see later on.

The development of the quadrupole glass in PMN is followed by the
transformation of the phonon dynamics to the order-disorder one. Indeed, we assume that the dipoles in the PNR's, due to the strong dipole-dipole interaction inside each of the PNR's, are relaxators. The growth of the correlation radius of the PNR's
can be associated with the growth of the relaxation
contribution to the dielectric permittivity on cooling. Simultaneously, when these
dynamical PNR's touch each other, i.e., in other words, start to acquire a static component of strain (this, we believe, happens
below $T_1^*$ \cite{Dulkin}), they organize a new phase with the
quadrupole order, which, fully or partially replaces the spherical glass state below $T_2^*$. Thus, we associate the temperature interval
[$T_2^*$,$T_1^*$] with the one, where the inhomogeneous strain organizes a steady state, and
the relaxation contribution to the dielectric permittivity becomes
critically important and practically substitutes the phonon contribution.
This consideration correlates well with the development of the dielectric
spectrum of PMN recently found experimentally by Bovtun \textit{et al} \cite{Bovtun}.

The quadrupole glass can be described phenomenologycally by employing the following simple free energy
\begin{eqnarray}
&F=\frac{1}{2}a_P(T-T_2^*)P^2+\frac{1}{4}\beta P^4 + \nonumber \\ &\frac{1}{2}a_Q(T-T_Q)Q^2+\frac{1}{3}bQ^3-e^2Q+\frac{1}{2}cP^2Q-e^2Q
\end{eqnarray}
where $P$ is polarization, Q is the quadrupole glass order parameter, $T_Q$ is the critical temperature, which, in the first approximation, equals $T_2^*$, $e^2$ is the random strain variance, which becomes significant below $T_1^*$. The equilibrium condition for $Q$ now reads
\begin{equation}
\label{Q}
a_Q(T-T_Q)Q+bQ^2-e^2+cP^2Q=0
\end{equation}
Solutions of Eq. (\ref{Q}) were studied in Ref. [\onlinecite{First}]. Here, we consider the case of large $e^2$:
\begin{equation}
\label{formula}
Q=\frac{-a(T-T_Q)+\sqrt{a^2(T-T_Q)^2+4b e^2}}{2b}
\end{equation}
This order parameter appears well above $T_Q$ and nearly linearly increases on cooling below $T=T_Q$.

The emergence of the quardrupole glass order parameter is also behind the suppression of the ferroelectric phase transition. The soft ferroelectric mode behavior can be described by the following simple formula
\begin{equation}\label{eq6}
\omega^2=\omega_0^2+\alpha (T-T_2^*) + c Q
\end{equation}
If we define temperature $T^*$ such that
\begin{equation}
\frac{\partial \omega^2 (T^*)}{\partial T} = 0
\end{equation}
then, in the closest vicinity of $T^*$,
\begin{equation}
\label{diffuse}
\omega^2(T)=\omega^2(T^*)+\frac{(T-T^*)^2}{\delta^2}
\end{equation}
Equation (\ref{diffuse}) with $\delta$ being representative of the diffuseness of the transition, is one the main great features describing the dielectric behavior of relaxors, and now, we see that it might arise owing to the diffuse phase transition from the spherical glass to the quadrupole glass phase.

Figure 2 presents the result of our extraction of the $Q(T)$ temperature dependence from the experimentally measured dielectric permittivity at 100 kHz, in the same manner as it was done in Ref. [\onlinecite{Viehland-Curie}]. The main result of this extraction is the following. $Q(T)$ is zero, above $T_d$. It appears at $T=T_d$, and gradually increases on cooling, down to $T_c=T_Q=385\,\rm{K}$. This latter temperature is the Curie-Weiss extrapolated critical temperature, from $T>T_d$ part. The smooth temperature behavior of $Q(T)$ is characteristic of diffuse phase transitions. Below $T_Q$, the dependence of $Q(T)$ becomes more and more linear that is inherent to normal behavior of a scalar order parameter like the quadrupole glass order parameter, in the Landau theory.

We tried fitting eq. (\ref{formula}) to the extracted from experiment $Q(T)$ curve, at a fixed value of $T_Q=385\,\rm{K}$. One can see from Figure 3 (curve 1) that the fitting curve well fits the linear portion of experiment, but lies well above experimental $Q(T)$, at high temperatures. This deflection naturally stems from the fact that the random strain variance $e^2$ in formula (\ref{formula}) is temperature independent. This makes $Q(T)$ finite at $T=T_d$ and above, in contrast to the experimental result. In order to fix this problem, one can follow the assumption about a temperature dependence of the random strain variance. Let $e^2=0$, at temperatures above $T=T_d$, and, at lower temperatures, $e^2$ gradually increases, until a maximal (saturated) magnitude of this variance. One of the ways to check this assumption is using a hyperbolic tangent: $e^2=e^2_0 tanh \frac{T_d-T}{2T_s}$ for $T<T_d$, and $e^2=0$ for $T>T_d$. Here $T_s$ is the temperature interval, where $e^2$ saturates. Figure 3 (curve 2) shows that this assumption well suits experiment. Thus, our data witness that the random field variance is zero, above $T=T_d$, starts from zero at Burns temperature and increases on cooling until some magnitude, and then saturates. Very probably, the saturation happens in the quadrupole phase.

\section{Diffuse scattering}
Figure 4 shows the integral intensity of the (001) Bragg peak extracted from neutron PMN powder diffraction pattern measured at Institut Laue-Langevin (Grenoble, France) between 100K and 850K. The temperature dependence of this intensity is representative for all the Bragg peaks (not shown here). Notice the changes of the slopes. The first change happens at a temperature located in the interval [$T_2^*$,$T_1^*$]. The second does at the so-called freezing temperature $T_f$ (see arrows). These changes are in good agreement with those observed on the temperature dependence of the lattice parameter (Figure 1). We assume that these abrupt changes of the integral intensity stem from the change of the diffuse scattering intensity. Indeed, it was shown earlier \cite{Vakhrushev-PMN,Gehring} that the diffuse scattering of neutrons with butterfly shape (see the inset in Figure 4) strongly increases below
$T_1^*$ on cooling, while some diffuse scattering exists even above $T_d$
\cite{Vakhrushev-prb} (the latter is known as Huang scattering on the quenched lattice parameter fluctuations, which can be described by chemical disorder of the Nb and Mg ions in the same sublattice). Notice that the nature of the butterfly diffuse scattering has been intensively discussed in literature (see the most resent discussion in Ref. [\onlinecite{Vakhrushev-PMN}] and references therein). This scattering, as it was understood, arises due to the thermal excitations of the combined optical and acoustic atomic vibrations, having strong anisotropy. The nature of this anisotropy is still debating. Thus, this scattering manifests, in some sense, the transformation of the spherical glass into an anisotropic glass (like the transformation of the Heisenberg spins into Ising ones \cite{Brahim,Stock,Jirka}).

Figure 5 presents fitting of the same function as in Fig. 4 (curve 2), differing by only a factor, to the integral intensity of the neutron data, after subtraction of the high-temperature value of this intensity \cite{footnote3}. One can see that this function nicely fits these data in the temperature interval above $T_f=200\,\rm{K}$.  This finding gives us significant information that the integral intensity of the diffuse scattering can be considered as a general order parameter of relaxors, in the same way as we got this parameter, $Q(T)$, above, from the dielectric permittivity data.

\section{Diffuse dielectric permittivity in a broad temperature range}

As it follows from equation (\ref{eq6}), the soft mode hardens, because of the appearance of $Q$. On the other hand, the appearance of $Q$ leads to a new contribution to the dielectric permittivity, due to the relaxations. The relaxation polarization, due to similarity in symmetry, may couple to the phonon displacements, and finally, can get cooperative. The combined phonon-relaxation response can be described analytically, in the mean-field approximation \cite{Vug-dyn,Vug-Rab,JETP-Trepakov}, and the final formula is given by equation (\ref{epsilon}). In the case if one considers zero-frequency, i.e. the dc (Field-Cooled) response, the trend of the total dielectric susceptibility, which includes both the relaxation and phonon parts is such that, below $T_Q \approx T_2^*$ the dielectric permittivity keeps growing \cite{Kutnjak}, with temperature decrease, and saturates, below $T_f$. This makes the quadrupole phase having an extremely large value of the dielectric permittivity, in a wide temperature interval. However, at finite frequencies, this permittivity strongly decreases, below a frequency-related temperature, owing to the Vogel-Fulcher temperature dependence of the relaxation time.

Thus, the colossal dielectric response in PMN arises from the combination of the minimum of the soft mode and critically increasing, on cooling, Edwards-Anderson parameter $Q$ corresponding to the quadrupole phase. These anomalies arise, in our opinion, due to three consequent diffuse phase transitions, at $T_d$, $T^*$, and $T_f$, first, owing to phonon softening, towards $T^*$, on cooling, and, then, due to the conversion of the phonon dynamics to the relaxation one, at lower temperatures, down to $T_f$. It is important to stress the cooperative character of the polarization dynamics, due to the coupling of the phonons with the relaxations.

Recent experimental data \cite{twoMode,Al-Zein} showed that, in PMN, there are actually two soft modes, $\omega_1$ and $\omega_{2}$, with quite different frequencies. This finding has not got yet a reasonable explanation. Let us suggest that these two modes correspond to the regions covered by the PNR's correlation spheres, in which the soft mode frequency $\omega_1$ is supposed to be high, due to the existence of the local $Q$ order parameter, and to the regions away from the PNR's volume, where the soft mode $\omega_2$ is not disturbed by $Q$. According to the experimental data \cite{twoMode}, the temperature dependence of  the mode $\omega_2$ frequency can be described by the Curie-Weiss law, with $T_c\approx 400\,\rm{K}$. This fact well corresponds to our assumption that $T_2^*\approx T_c$, and that this temperature corresponds to the phase transition from the spherical glass to quadrupole glass phase. The temperature dependence of $\omega_1$ is much weaker \cite{twoMode}, and this also corresponds to our assumption that the softening of this mode is responsible for the phase transition from the quadrupole to ferroglass phase. In order to check our assumption, one can measure the temperature dependency of the relative weight of these two modes. We suppose that $\omega_1$ becomes more populated than $\omega_1$, with temperature decrease, because of the development of the quadrupole phase on cooling.

One more remark concerns our assumption about the closeness of $T_c$ and $T_Q^*$ (see above). This conception assumes that, above $T_d$, PMN presents a well seen development of the ferroelectric frequency softening, Ornstein-Cernice (Landau) correlations, and Curie-Weiss behaviour of the dielectric permittivity. However, at lower temperatures, another tendency competes with the ferroelectric scenario. We believe that the appearance of $Q$ order parameter can be described in a manner similar to the one suggested in Ref. [\onlinecite{Tagantsev}] (for the first general mathematical exploration if this competition see Ref. [\onlinecite{Holakovsky}], which introduced a new, trigger type, phase transition). These authors suggested that, in some of the relaxors, there can exist a scenario of a local phase transition and heterophase fluctuations. More specifically, the nonpolar local phase transition can happen in, say, PMN, at $T_d$, because of the competition between the ferroelectric and, in our terminology, spherical glass order parameter. Such a transition must be of first order, but, in our opinion, might be diffuse \cite{diffuse}. So, similarly, we suggest that such kind of the phase transformation may happen also at a lower temperature, which we denoted as $T_Q^*$. Here the phase transformation occurs between the spherical glass and quadrupole glass phases, and this transition, as we showed above (see also Ref. [\onlinecite{Viehland}]) to be strongly diffuse. Interestingly, at a lower temperature, $T=T_f$, we again meet absolutely the same situation, but, now, between the quadrupole- and ferro- glass phases \cite{First}. Thus, in our opinion, all three transformations follow the Chu-Setter-Tagantsev scenario \cite{Tagantsev} suggested earlier also as a general trigger mechanism [\onlinecite{Holakovsky}].

\section{Discussion of the possible tetragonal distortions in PMN}

In this section, we will discuss possible scenarios of the tetragonal deformation of the lattice of PMN revealed by the Raman studies \cite{Svitelskiy}. First of all, as soon as the shape of the crystal, in the whole temperature range, remains cubic, these tetragonal distortions must be local. In principle, the source for this distortion can be whether the strong dipole-dipole interactions between nearest Pb displacements or the elastic (or antiferroelectric) interaction between next-to-nearest-neighbors. In the former case, the Pb pairs, having in PMN [001] directions, can serve as tetragonal seeds for the lattice. In the latter case, the disordered incommensurate patterns, for example, in [110] directions, can result in the local tetragonal distortions. In general, the competition between the ferroelectric and antiferroelectric order can be behind the emergence of the stable PNR's with low local symmetry \cite{Laurent-BZT,ProsandeevFTT,Tkachuk,Sobolev,Gehring-boundary,domains,Miao}. As a consequence, these nano-objects can break the macroscopic polarization and help the emergence of
the quadrupole glass phase. 

As an example of such glass formation, the Li pairs in the solid solution KTaO$_3$:Li (KLT) are tetragonal centers and organize an orthogonal (90-degree) patterns made of these pairs, owing to the next-to-nearest elastic interactions \cite{KLT}. The situation in PMN can be different, because the Pb sublattice does not contain any other ions but lead (KLT contains K and Li in the same sublattice). However, the random fields, created by the random distribution of the Mg and Nb, in PMN, can make the lead ions different, and this difference is characterised by the magnitude (and, generally, direction) of the local electric field. For example, it was realized earlier that those lead ions, which possess the smallest local electric field, stemming from the random distribution of the Mg and Nb ions, are more ferroelectrically active, while those, which are embedded into a strong local field, are not ferroelectrically active \cite{Bokov}. 
It is worth citing the recent first-principles calculations of the PMN relaxors diluted with PbTiO$_3$ revealing the tetragonal prevalence in the correlations between the atomic displacements \cite{Rappe}.

Weak crystallization theory might be an adequate model for the PNR's growth in PMN \cite{Kats}. According to this theory, the nuclei of the new phase grow in such a manner that the magnitude of the correlation radius is an order parameter, but the direction is not (the symmetry of the nucleus is spherical). A similar situation was recently shown to have some relevance to Higgs boson \cite{Higgs}. Another example is the labyrinth structures observed in ultrathin films of lead titanate \cite{Streiffer}.

 To answer the question about the reasons resulting in the tetragonal splitting of the Raman lines in PMN \cite{Svitelskiy}, one needs performing more additional experiments and derivations.

\section{Photoelectric current}
We have also considered the effect of the dynamic
polarization, inherent to the quadrupole phase, on the dc photoconductivity
$\sigma_{ph}$ of PMN. We studied a flux-grown PMN crystal having
a form of cube (with edges of 1.8 mm) and with the faces parallel to the
(001) planes of the prototype perovskite lattice. We put electrodes made of
Aquadag on two opposite faces of the crystal parallel to the incident
light beam. We illuminated the crystal with incandescent light having the
intensity of 100 mW. The absence of substantial heating of the crystal by
the incident light was proved by close proximity of the permittivity-temperature
curves measured at 1 kHz, in the darkness and under continuous illumination.
Electric current was measured during slow (2-3 K/min) heating rate by
means of a V7-30 electrometer. Figure 6 presents the results obtained. At low
temperatures, below 210 K i.e. $T_f$, we found the logarithm of the photoelectric
current increasing approximately linearly with respect to inverse
temperature, like in semiconductors, with an activation energy of 0.08 eV.
Between 210 K and approximately 240 K, the photoelectric current
heavily decreased manifesting a catastrophic decrease of the mobility (this
finding is in line with the earlier measurements of the electron mobility in
PMN \cite{Trepakov}), which we relate to the scattering of the electrons by the
fluctuations of the crystal structure at the border between the ferro- and
quadrupole glasses. Then above 240 K, our experiment again shows a
linear dependence of the logarithm of the electric current on the inverse
temperature, but, this time, with a much larger activation
energy than at low temperatures i.e. 0.35 eV  against 0.08 eV at low temperature. This very strong increase of the activation energy is in
line with our assumption of the existence in the lattice of a strong
(relaxation) dynamic of the polarization. This dynamic makes the dielectric
permittivity having extremely large magnitudes, and is reflected in the
large magnitudes of the electric conductivity activation energies inherent
to polarons \cite{Raevski1998}. It is worth noting that similar decrease of $\sigma_{ph}(T)$ on heating was observed previously in the region of
phase transitions between the ferroelectric and antiferroelectric phases in
NaNbO$_{3}$ \cite{Raevski1979} and between ferroelectric and
paraelectric phases in Cd$_{2}$Nb$_{2}$O$_{7}$
\cite{Raevski1983}. It should be stressed that the measuring field strength was
rather small ($\approx 100\,\rm{V/cm}$ along the [001] direction) and could not
induce the ferroelectric phase at low temperatures.

\section{Summary}
Finally, we have developed a model of PMN relaxor involving the recently discovered $T^*$ temperature. Our model considers $T^*$ as a quasicritical temperature which manifests the transformation of the lead displacement pattern from the
spherical glass to a quadrupole glass order. One of the
evidences of the anisotropic order is the intense butterfly-like diffuse
scattering which develops in this temperature
interval, the tetragonal splitting of the Raman
lines, specific features in the neutron scattering \cite{Stock}, and the Viehland \textit{et al} \cite{Viehland} evidence of the glass order parameter. We believe that, on further temperature decrease, the quadrupole glass transforms
into a ferroglass phase. We argued that all three diffuse phase transitions (transformations) (at $T_d$, $T_Q$, and $T_f$) follow the Chu-Setter-Tagantsev scenario known generally also as a trigger mechanism \cite{Holakovsky}. We have emphasized that the quadrupole glass possesses a cooperative relaxation-phonon dynamics, looking as cooperative beatings of
the local polarization, which have some similarities with the
boson vortex dynamics in the quantum phase of $^4$He and with the zero-point atomic vibrations, in quantum paraelectrics. These beatings (cooperative relaxations) suppress
the development of the lattice parameter, form a colossal dielectric
permittivity response, and result in a polaronic-like effect for the
electrons.  We hope that our model will find further
experimental evidences, theoretical exploration and will be utilized to take advantage of the  discovered transformations in the applications desiring giant dielectric responses, of non-electron nature, in wide temperature intervals around room temperature.

\section{Acknowledgement}
SP and BD appreciate discussions of the topic with Sergey Vakhrushev, Alexander Tagantsev, and Peter Gehring.
This study is partially supported by Russian Foundation for Basic Research
Grant No. 12-08-00887-a. SP also
appreciates Office of Naval Research Grants N00014-12-1-1034 and
N00014-11-1-0384.

\begin{figure}
\centering
\resizebox{0.9\textwidth}{!}
{\includegraphics{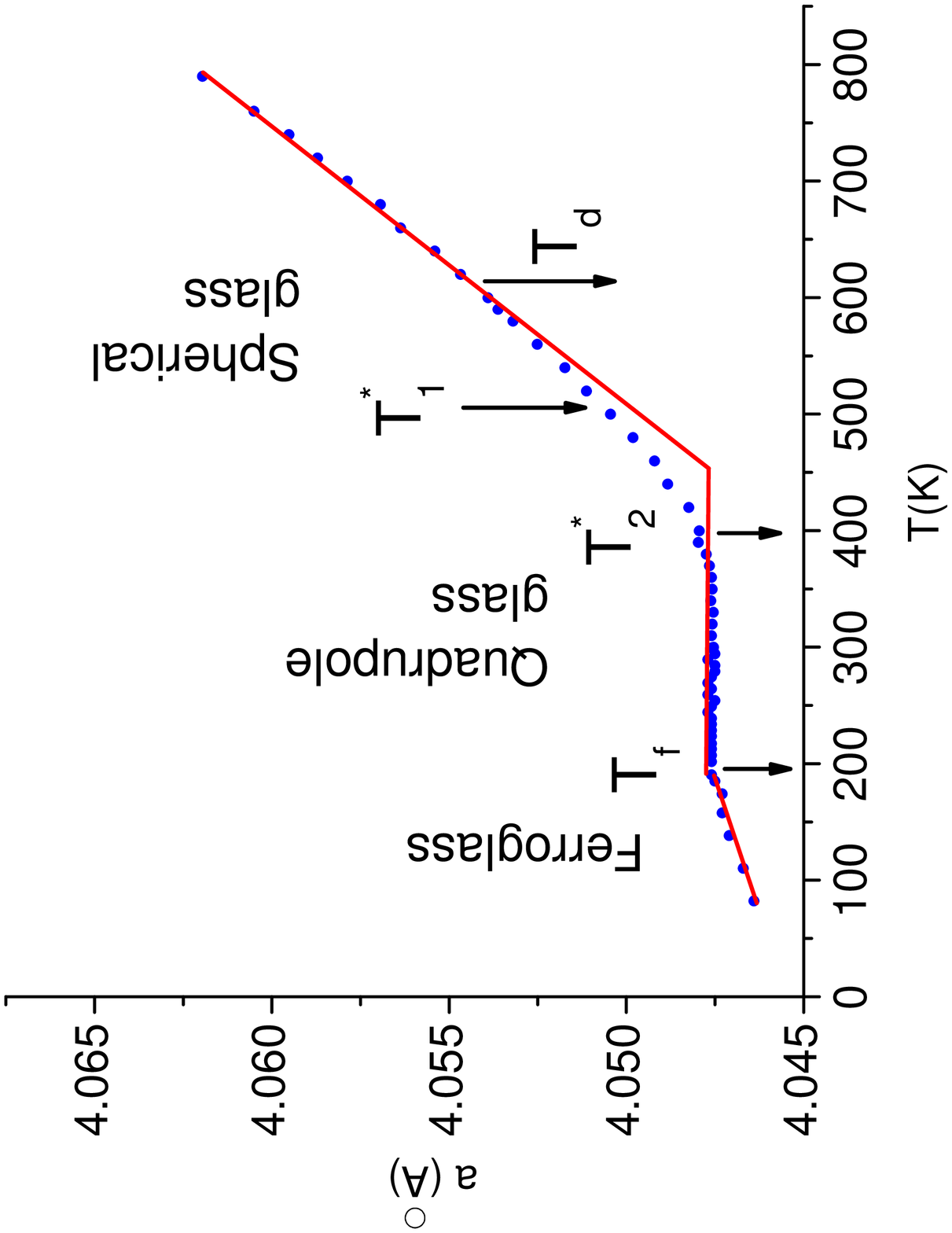}}
\caption{Pseudocubic lattice parameter $a$ as a function of temperature obtained with the (200), (220), and (222) Bragg peaks for PMN. The straight lines are guide to the eyes.} \label{F1}
\end{figure}

\begin{figure}
\centering
\resizebox{0.9\textwidth}{!}{\includegraphics{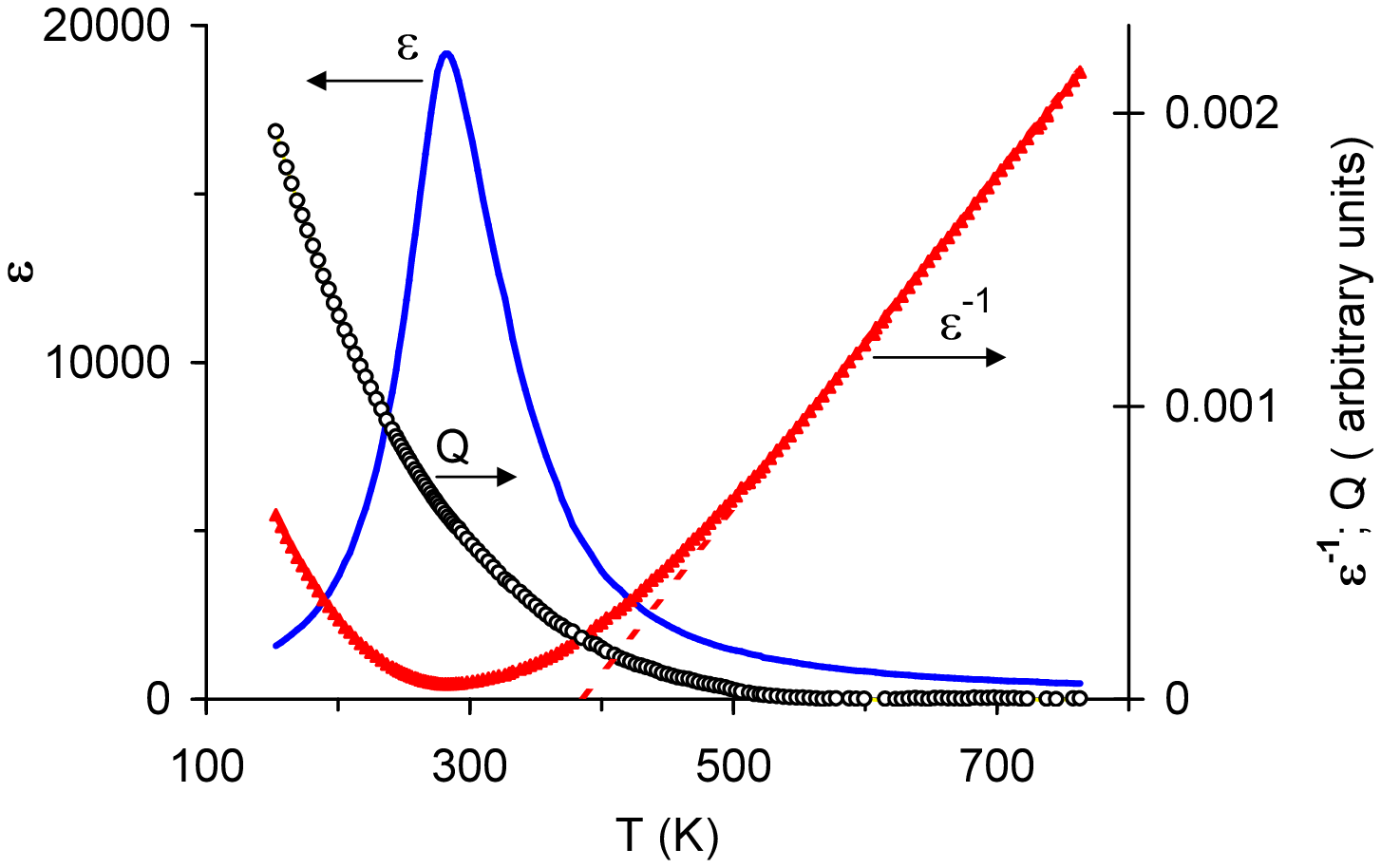}}
\caption{ Temperature dependence of dielectric permittivity $\varepsilon$ measured at 100 kHz, inverse permittivity $\varepsilon^{-1}$ and glass order parameter $Q$ for (001)-oriented PMN single crystal.
} \label{F4}
\end{figure}

\begin{figure}
\centering
\resizebox{0.9\textwidth}{!}{\includegraphics{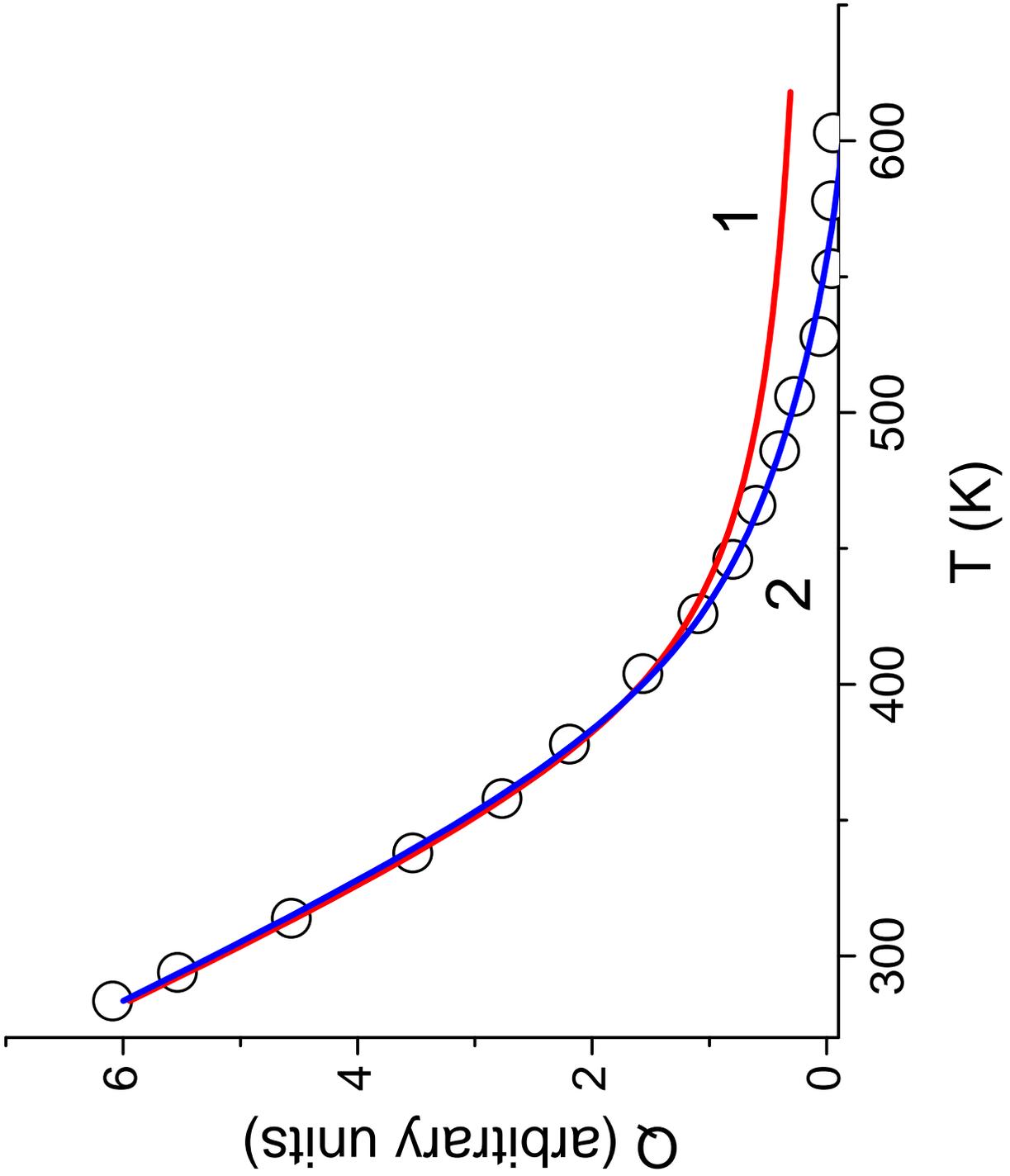}}
\caption{ Fitting formula (5), with $\delta=\delta_0 coth \frac{T_d-T}{2T_s}$ for $T<T_d$ and $\delta=0$  for  $T>T_d$, to the experimental curve shown in Fig. 2, after subtraction the high-temperature constant background.
} \label{F7}
\end{figure}

\begin{figure}
\centering
\resizebox{0.9\textwidth}{!}{\includegraphics{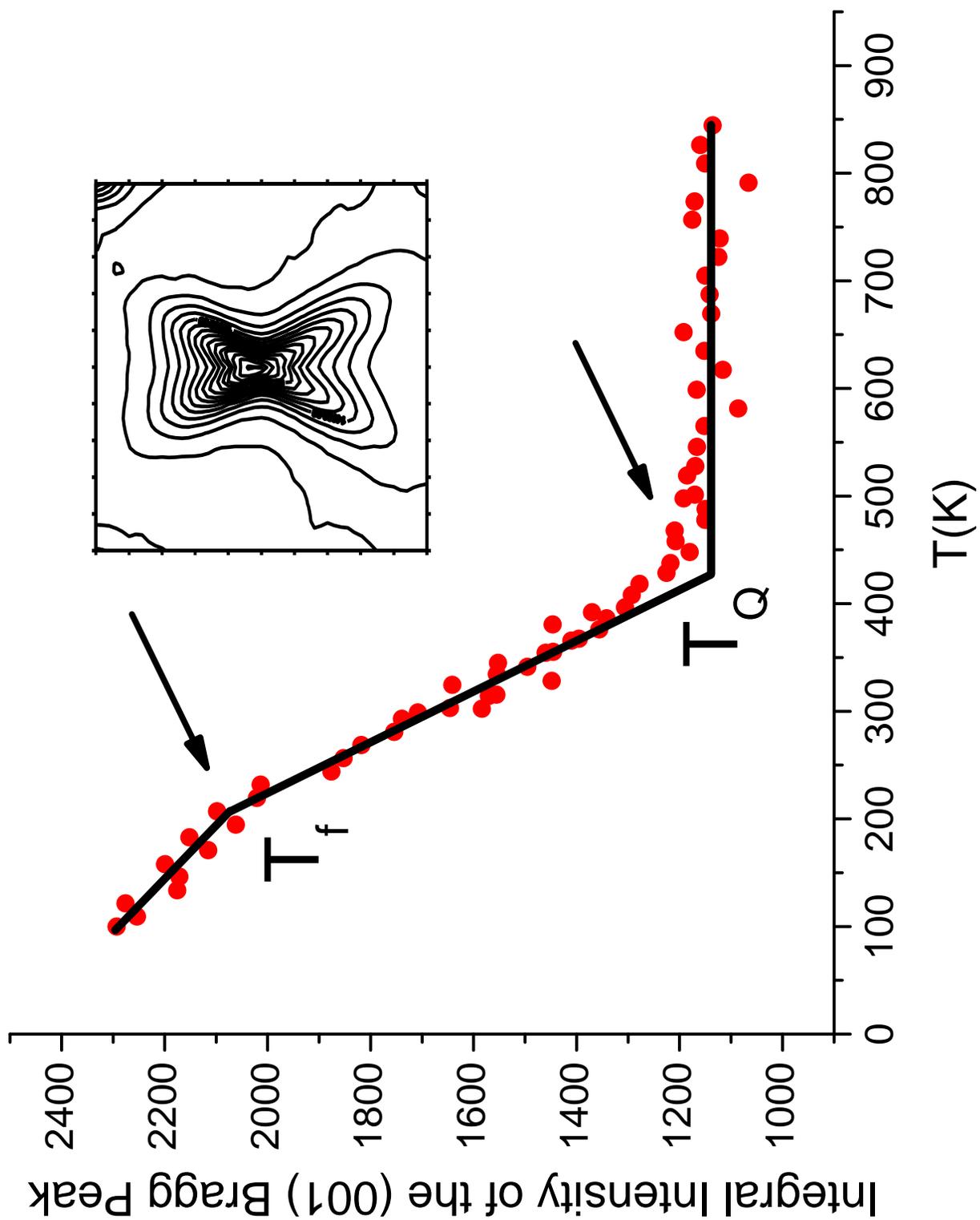}}
\caption{ Temperature dependence of the integral intensity of the (001) Bragg peak extracted from the neutron powder diffraction pattern in PMN. The inset shows the experimentally measured butterfly shape of the diffuse scattering.
} \label{F6}
\end{figure}

\begin{figure}
\centering
\resizebox{0.9\textwidth}{!}{\includegraphics{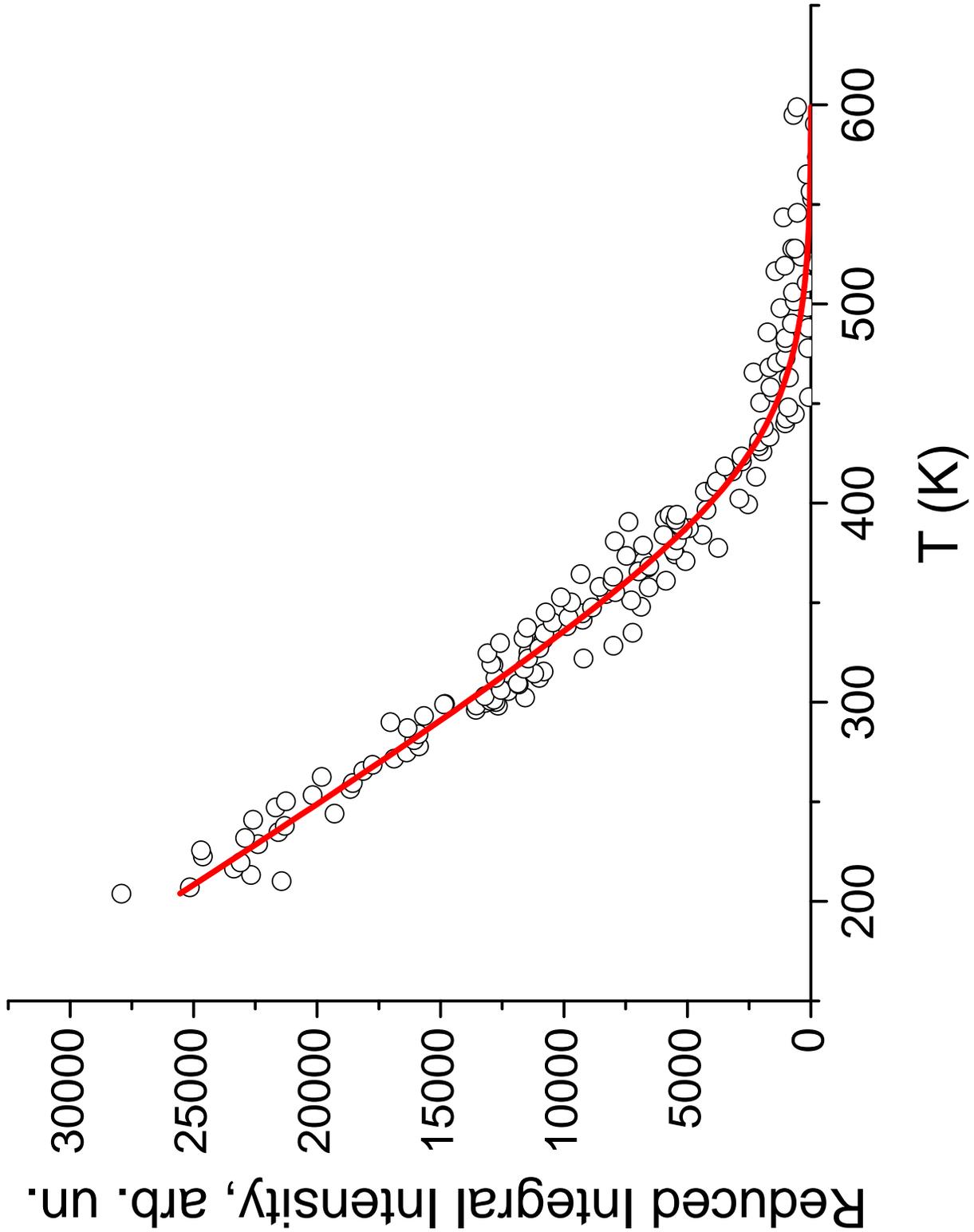}}
\caption{ Fitting equation (5) to the temperature dependence of the glass order parameter $Q$, for (001)-oriented PMN single crystal, using $T_Q=385\,\rm{K}$ (1) and using the assumption that $e^2=e^2_0 tanh \frac{T_d-T}{2T_s}$ for $T<T_d$ and $e^2=0$  for  $T>T_d$ (2) (The latter assumption implies that $T=T_d$ is a critical temperature, for the appearance of dilatation centers, or, in other words, spherical lead off-centers).
} \label{F5}
\end{figure}



\begin{figure}
\centering
\resizebox{0.8\textwidth}{!}{\includegraphics{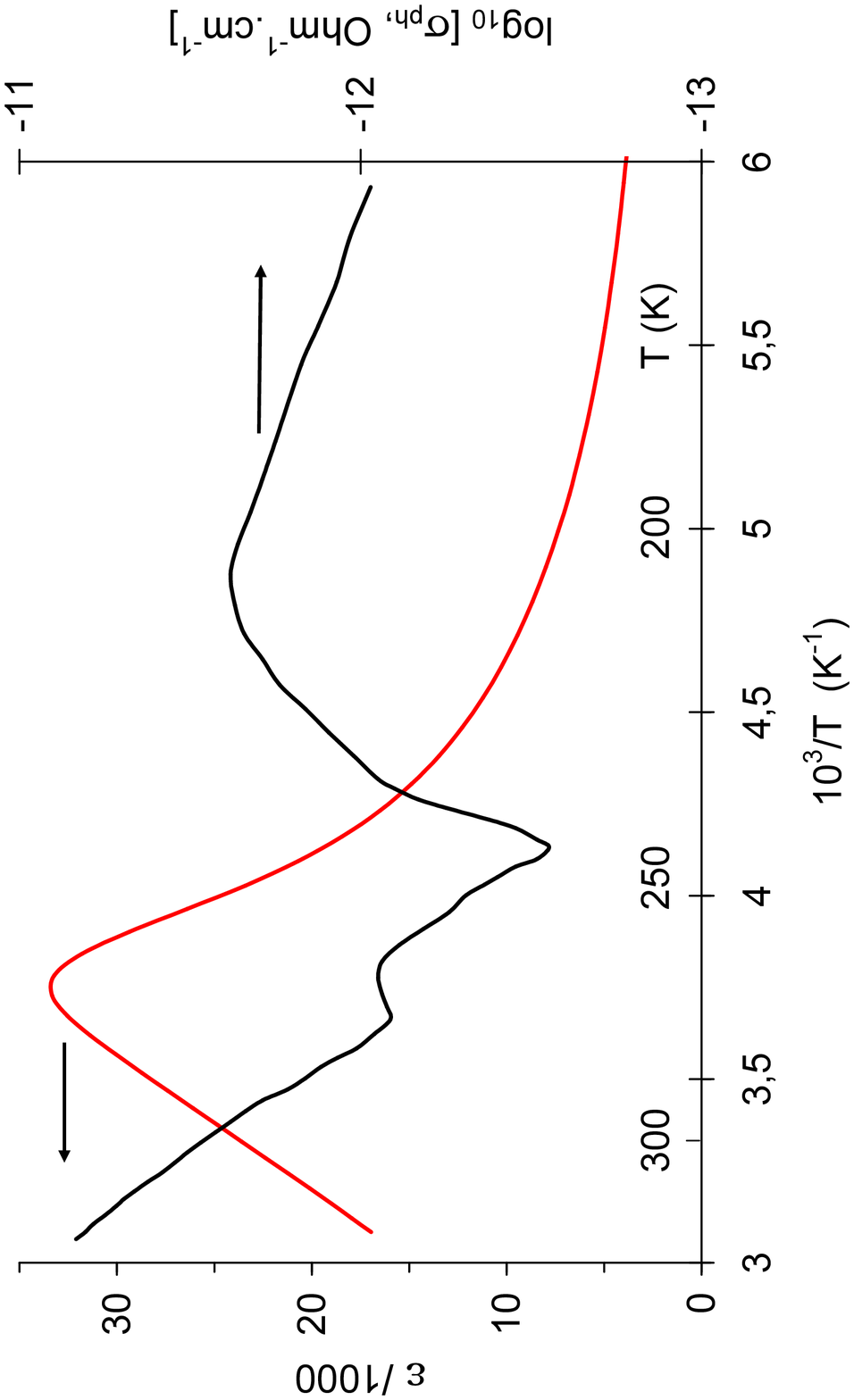}}
\caption{ Temperature dependency of dc photoconductivity and ac dielectric permittivity (at 1 kHz), for (001)-oriented PMN single crystal.
} \label{F8}
\end{figure}


\begin{thebibliography}{1900}
\bibitem{Smolenskiy}G. A. Smolensky and A. I. Agranovskaya, Sov. Phys. Tech. Phys.
\textbf{3}, 1380 (1958).
\bibitem{Cross}L. E. Cross, Ferroelectrics \textbf{151}, 305 (1994).
\bibitem{Viehland}D. Viehland, S. J. Jang, L. E. Cross, and M. Wuttig, Philos. Mag. B \textbf{64}, 335 (1991).
\bibitem{Zhang}S. Zhang and F. Li J. Appl. Phys. \textbf{111}, 031301 (2012).
\bibitem{Glazunov}A. E. Glazounov, A. J. Bell, and A. K. Tagantsev, J. Phys.:
Condens. Matter \textbf{7}, 4145 (1995).
\bibitem{Isupov}V. A. Isupov. Phys. Stat. Sol. (b) \textbf{213}, 211 (1999); V. A. Isupov, Sov. Phys. Solid State \textbf{28} (1986) 1253.
\bibitem{Colla}E. V. Colla, E. Yu. Koroleva, N. M. Okuneva, and S. B. Vakhrushev, Phys. Rev. Lett. \textbf{74}, 1681 (1995).
\bibitem{Vugmeister-Glinchuk}B. E. Vugmeister and M. D. Glinchuk, Rev. Mod. Phys. \textbf{82} 993 (1990).
\bibitem{Blinc1}R. Pirc and R. Blinc, Phys. Rev. B \textbf{60}, 13470 (1999).
\bibitem{Blinc2}R. Blinc, J. Dolin\v{s}ek, A. Gregorovi\v{c}, B. Zalar, C. Filipi\v{c}, Z. Kutnjak, A. Levstik, and R. Pirc, Phys. Rev. Lett. \textbf{83} 424 (1999).
\bibitem{Blinc3}R. Pirc, R. Blinc, and V. Bobnar, Phys. Rev. B \textbf{63}, 054203 (2001).
\bibitem{Kalinin}A. L. Kholkin , A. N. Morozovska, D.
A. Kiselev, I. K. Bdikin, B. J. Rodriguez, P. Wu, A. A. Bokov, Z.-G. Ye, B. Dkhil, L.-Q. Chen, M.
Kosec, S. V. Kalinin, Adv. Func. Mat. \textbf{11}, 1977 (2011).
\bibitem{Cowley}R. A. Cowley, S. N. Gvasaliya, S. G. Lushnikov, B. Roessli,
and G. M. Rotaru, Advances in Physics \textbf{60}, 229 (2011).
\bibitem{Bokov}A. A. Bokov and Z. G. Ye, Phys. Rev. B \textbf{66}, 064103 (2002).
\bibitem{Burns}G. Burns and F. H. Dacol, Solid State Commun. \textbf{48}, 853 (1983).
\bibitem{Tagantsev}F. Chu, N. Setter, and A. K. Tagantsev, J. Appl. Phys. \textbf{93}, 5129 (1993).
\bibitem{Okuneva}S.B. Vakhrushev, and N.M. Okuneva, AIP Conf. Proc. \textbf{626}, 117 (2002).
\bibitem{Vakhrushev}S. B. Vakhrushev, A. A. Naberezhnov, N. M. Okuneva, and B.
N. Savenko, Phys. Solid State \textbf{37}, 3621 (1995);
S. B. Vakhrushev, B. E. Kvyatkovsky, A. A. Naberezhnov, N. M.
Okuneva, and B. Toperverg, Ferroelectrics \textbf{90}, 173 (1989).
\bibitem{Jirka}M. Pa\'{s}ciak, T. R. Welberry, J. Kulda, M. Kempa, and J. Hlinka, Phys. Rev. B \textbf{85}, 224109 (2012).
\bibitem{Stock}C. Stock, Guangyong Xu, P. M. Gehring, H. Luo, X. Zhao, H. Cao, J. F. Li, D. Viehland, and G. Shirane, Phys. Rev. B \textbf{76}, 064122 (2007).
\bibitem{Bussmann}A. R. Bishop, A. Bussmann-Holder, S. Kamba, and M. Maglione,  Phys. Rev. B \textbf{81}, 064106 (2010).
\bibitem{Higgs}D. Podolsky, A. Auerbach, and D. P. Arovas, Phys. Rev. B \textbf{84}, 174522 (2011).
\bibitem{Langevin}S. A. Prosandeev, Phys. Sol. State \textbf{45}, 1774 (2003).
\bibitem{nonlinear}S. A. Prosandeev, I. P. Raevski, A. S. Emelyanov, \textit{et al},   J. Appl. Phys. \textbf{98},   014103    (2005).
\bibitem{Tagantsev-Vogel}A. K. Tagantsev, Phys. Rev. Lett. \textbf{72} 1100 (1994).
\bibitem{RandomField}V. Westphal, W. Kleemann, and M. D. Glinchuk, Phys. Rev. Lett. \textbf{68}, 847 (1992).
\bibitem{domains}S. A. Prosandeev, U. Waghmare, I. P. Raevski and L. Jastrabik, Integrated Ferroelectrics \textbf{58}, 1359 (2003); S. A. Prosandeev, I. P. Raevski and U. V. Waghmare, AIP Conference Proceedings,677, 41 (2003); arXiv:cond-mat/0302496.
\bibitem{Glinchuk} M. D. Glinchuk, British Ceramic Transactions \textbf{103}, 76 (2004).
\bibitem{Egami}W. Dmowski, S. B. Vakhrushev, I.-K. Jeong, M. P. Hehlen, F. Trouw, and T. Egami, Phys. Rev. Lett. \textbf{100}, 137602 (2008). I.-K. Jeong,T. W. Darling, J. K. Lee, Th. Proffen, R. H. Heffner, J. S. Park, K. S. Hong, W. Dmowski, and T. Egami, Phys. Rev. Lett. \textbf{94}, 147602 (2005).
\bibitem{Glinchuk-Stephanovich}M. D. Glinchuk and V. A. Stephanovich, J. Phys.: Condensed Matter \textbf{6}, 6317 (1994).
\bibitem{Barton}S. Tinte, B. P. Burton, E. Cockayne, and U.V. Waghmare,
Phys. Rev. Lett. \textbf{97}, 137601 (2006).
\bibitem{Akbas}P. K. Davies and M. A. Akbas, J. Phys. Chem. Solids \textbf{61},
159 (2000).
\bibitem{Laurent-BZT}A. R. Akbarzadeh, S. Prosandeev, Eric J. Walter, A. Al-Barakaty, and L. Bellaiche, Phys. Rev. Lett. \textbf{108}, 257601 (2012).
\bibitem{Iniguez}Jorge \'I\~niguez and L. Bellaiche, PHYSICAL REVIEW B \textbf{73}, 144109 (2006).
\bibitem{Rappe}Hiroyuki Takenaka, Ilya Grinberg, and Andrew M. Rappe, Phys. Rev. Lett. \textbf{110}, 147602 (2013).
\bibitem{Nakhamson}S. M. Nakhmanson and I. Naumov, Phys. Rev. Lett. \textbf{104}, 097601 (2010).
\bibitem{Narayani}N. Choudhury, L. Walizer, S. Lisenkov and L. Bellaiche, Nature \textbf{470}, 513 (2011).
\bibitem{Viehland-Curie}D. Viehland, S. J. Jang, L. E. Cross,and M. Wuttig, Phys. Rev. B \textbf{46}, 8003(1992).
\bibitem{Brahim}B. Dkhil, J. M. Kiat, G. Calvarin, G. Baldinozzi, S. B. Vakhrushev,
and E. Suard, Phys. Rev. B \textbf{65}, 024104 (2001).
\bibitem{Svitelskiy}O. Svitelskiy, J. Toulouse, G. Yong, and Z.-G. Ye, Phys. Rev. B \textbf{68}, 104107 (2003).
\bibitem{Dulkin}B. Dkhil, P. Gemeiner, A. Al-Barakaty, L. Bellaiche, E. Dulkin,
E. Mojaev, and M. Roth, Phys. Rev. B \textbf{80}, 064103 (2009).
\bibitem{Gehring}C. Stock, L. Van Eijck, P. Fouquet, M. Maccarini, P. M. Gehring, Guangyong Xu, H. Luo, X. Zhao, J.-F. Li, and D. Viehland, Phys. Rev. B \textbf{81}, 144127 (2010).
\bibitem{footnote1}The higher precision means here smaller wave vectors, and, correspondingly, larger affected regions.
\bibitem{Ivliev}M.P. Ivliev, V.P. Sakhnenko,   Soviet Phys. Solid State Phys. \textbf{28}, 356 (1986).
\bibitem{Vollmayr}H. Vollmayr, R. Kree, and A. Zippelius, Phys. Rev. B \textbf{44}, 12238 (1991).
\bibitem{Michel}R. M. Lynden-Bell and K. H. Michel, Rev. Mod. Phys. \textbf{66}, 721 (1994).
\bibitem{Pirc}B. Tadic, R. Pirc, and R. Blinc, Phys. Rev. B \textbf{55}, 816 (1997).
\bibitem{footnote2}see similar results, but in a smaller temperature interval, in Ref. [\onlinecite{Dulkin}].
\bibitem{Landau}L. D. Landau and E. M. Lifshitz, Statistical Physics
(Butterworth-Heinemann, London, 1980), Vol. 5, Part 1.
\bibitem{Dulkin09} E. Dul'kin, E. Mojaev, M. Roth, I. P. Raevski, and S. A. Prosandeev, Appl. Phys. Lett. \textbf{94}, 252904 (2009).
\bibitem{BZT-a} Ph. Sciau and A-M. Castagnos, Ferroelectrics \textbf{270}, 259 (2001).
\bibitem{Tipler}F. London, Superfluids (New York: Dover Publications, Inc., 1964).
P. A. Tipler and R. A. Llewellyn, {\it Modern Physics}, 6-th eddition, (W.H. Freeman and Co, New York 2012).
\bibitem{Muller}K. A. M\"{u}ller and H. Burkard, Phys. Rev. B \textbf{19}, 3593 (1979).
\bibitem{Vug-Rab}B. E. Vugmeister and H. Rabitz, Phys. Rev. B \textbf{57} 7581 (1998).
\bibitem{Kutnjak}A. Levstik, Z. Kutnjak, C. Filipi\v{c}, and R. Pirc, Phys. Rev. B \textbf{57}, 11204 (1998).
\bibitem{Pirc-ECE}R. Pirc, Z. Kutnjak, R. Blinc, and Q. M. Zhang, J. Appl. Phys. \textbf{110}, 074113 (2011).
\bibitem{Vug-glass}B. E. Vugmeister and H. Rabitz, J. Phys. Chem. Sol. \textbf{61}, 261 (2000).
\bibitem{Blinc4}R. Pirc, R. Blinc, V. Bobnar, and A. Gregorovi\v{c}, Phys. Rev. B \textbf{72}, 014202
(2005).
\bibitem{Binder}K. Binder and A. P. Young, Rev. Mod. Phys. \textbf{58}, 801 (1986).
\bibitem{E-A}S. F. Edwards and P.W. Anderson, J. Phys. F \textbf{5}, 965 (1975).
\bibitem{foot-vug}Note that, in the model of Ref. [\onlinecite{Vug-glass}], the glass order parameter does not have formal meaning of the glass order, but rather reflects the fact that the relaxors contain the relaxation processes, which might be longer than the inverse measuring frequency.

\bibitem{PMN}S. A. Prosandeev, E. Cockayne, B. P. Burton, et al. Phys. Rev. B  \textbf{70}, 134110   (2004).
\bibitem{KLT}S. A. Prosandeev, E. Cockayne, and B. P. Burton, Phys. Rev. B \textbf{68}, 014120 (2003).
\bibitem{Bovtun}V. Bovtun, S, Kamba, A. Pashkin, P. Samoukhina, J. Petzelt, I. P. Bykov, and M. D. Glinchuk, Ferroelectrics \textbf{298}, 23 (2004). V. Bovtun, J. Petzelt, V. Porokhonskyy, S.Kamba, and Y. Yakimenko. J. Europ. Ceram. Soc. \textbf{21}, 1307(2001).
\bibitem{First}S. Prosandeev, M. Panchelyuga, S. Raevskaya, and I. Raevski, Appl. Phys. Lett. \textbf{91}, 242904 (2007).
\bibitem{Vakhrushev-PMN}A. Bosak, D. Chernyshov, Sergey Vakhrushev, and M. Krisch, Acta Cryst. A \textbf{68}, 117 (2012).
\bibitem{Vakhrushev-prb}R. G. Burkovsky, A. V. Filimonov, A. I. Rudskoy, K. Hirota and M. Matsuura, S. B. Vakhrushev, Phys. Rev. B \textbf{85}, 094108 (2012).
\bibitem{footnote3}We subtracted a constant from the experimental data, so, to have zero at high temperatures. This zero is the origin for our fit. One can see from Figure 4, where we present the original (untreated) data, that the subtracted constant is of about half of the total maximal intensity. In principle, Bragg intensity may vary with temperature by mild release of neutron extinction due to small local distortion, in case of relaxors. Thus, the tail in Figure 5 may contain the noncritical scattering, in particular, Huang scattering, by the imperfections. This part we describe by the random strain, and we show that the temperature dependent part of this scattering (we are not interested in the temperature independent part here) does not originate from the defects, but rather it comes from the PNR's emerging at $T=T_d$ (otherwise, the tail would go father than $T=T_d$). In addition, diffuse scattering also may vary by spectrometer condition. This part is also taken into account by the random strain tensor. Indeed, by changing the precision of the experiment, one simply changes the important wave vectors, that is the important scale, which, obviously, corresponds to the scale of the random strain tensor.
\bibitem{JETP-Trepakov}S. A. Prosandeev, and V. A. Trepakov, JETP, \textbf{94}, 419 (2002).
\bibitem{Vug-dyn}B. E. Vugmeister, Phys. Rev. B \textbf{73} 174117 (2006).

\bibitem{twoMode}S. B. Vakhrushev, P. G. Burkovskiy, S. Shapiro, and A. Ivanov, Phys. Solid State  \textbf{52}, 889 (2010).
\bibitem{Al-Zein}A. Al-Zein, J. Hlinka, J. Rouquette, and B. Hehlen, Phys. Rev. Lett. \textbf{105}, 017601 (2010).
\bibitem{Holakovsky}J. Holakovsky, Phys. Status Solidi B \textbf{56}, 615 (1973).
\bibitem{diffuse}I. P. Raevski, S. I. Raevskaya, S. A. Prosandeev, V. A. Shuvaeva,
A. M. Glazer, and M. S. Prosandeeva, J. Phys.: Condens. Matter \textbf{16}, L221 (2004).
\bibitem{ProsandeevFTT}S. A. Prosandeev, M. S. Panchelyuga, S. I. Raevskaya, and I. P. Raevskii., Phys. Solid State, \textbf{53}, 147 (2011).
\bibitem{Tkachuk}A. Tkachuk and H Chen, AIP Conf. Proc. \textbf{677},
55 (2003).
\bibitem{Sobolev}V. M. Ishchuk, V. N. Baumer, and V. L. Sobolev, J. Phys.
Condens. Matter \textbf{17}, L177 (2005).
\bibitem{Gehring-boundary}I. P. Swainson, C. Stock, P. M. Gehring, Guangyong Xu, K. Hirota, Y. Qiu, H. Luo, X. Zhao,
J.-F. Li, and D. Viehland, Phys. Rev. B \textbf{79}, 224301 (2009).
\bibitem{Miao}S. Miao, J. Zhu, X. Zhang, and Z.-Y. Cheng, Phys. Rev. B \textbf{65}, 052101 (2001).
\bibitem{Kats}E. I. Kats, V. V. Lebedev, and A. R. Muratov, Physics Reports \textbf{228}, 1 (1993).
\bibitem{Streiffer}D. D. Fong, G. B. Stephenson, S. K. Streiffer,
J. A. Eastman, O. Auciello, P. H. Fuoss,
C. Thompson, Science \textbf{304}, 1650 (2004).
\bibitem{Trepakov} V.A. Trepakov, A.V. Babinskii, N.N. Krainik, G.A. Smolenskii, and A.N. Samukhin, JETP Lett. \textbf{26},341(1977).
\bibitem{Raevski1998} I. P. Raevski, S. M. Maksimov, A. V. Fisenko, S. A. Prosandeev,J. Phys.: Condens. Matter \textbf{10}, 8015 (1998).
\bibitem{Raevski1979} I. P. Raevskii , M. A. Malitskaya, O. I. Prokopalo, V. G.
Smotrakov, and E. G. Fesenko, Soviet Physics - Solid State \textbf{21},  715 (1979).
\bibitem{Raevski1983} I. P. Raevski , M. A. Malitskaya, P. F. Tarasenko, O. I.
Prokopalo, Ya. E. Cherner, and V. E. Zakapko, Soviet
Physics - Solid State \textbf{25}, 1621 (1983).

\end{thebibliography}
\end{document}